\documentclass[runningheads]{llncs}
\usepackage{graphicx}
\usepackage{subfig}
\usepackage{hyperref}
\usepackage{academicons}
\usepackage{xcolor}
\usepackage{caption}
\usepackage{mathtools}
\usepackage{tabu}
\usepackage{multirow}
\usepackage{colortbl}
\def\filipolong{FiLiPo (\textbf{Fi}nding \textbf{Li}nkage \textbf{Po}ints)}
\def\filipo{FiLiPo}

\begin{document}

\title{FiLiPo: A Sample Driven Approach for Finding Linkage Points between RDF Data and APIs (Extended Version)}
\titlerunning{A Sample Driven Approach for Finding Linkage Points}
%
\author{Tobias Zeimetz\orcidID{0000-0002-5436-637X} \and
Ralf Schenkel\orcidID{0000-0001-5379-5191}}
\institute{Trier University, 54286 Trier, Germany\\
\email{\{zeimetz,schenkel\}@uni-trier.de}}

\maketitle

\begin{abstract}
  Data integration is an important task in order to create comprehensive RDF knowledge bases. Many data sources are used to extend a given dataset or to correct errors. Since several data providers make their data publicly available only via Web APIs they also must be included in the integration process. However, APIs often come with limitations in terms of access frequencies and speed due to latencies and other constraints. On the other hand, APIs always provide access to the latest data. So far, integrating APIs has been mainly a manual task due to the heterogeneity of API responses. To tackle this problem we present in this paper the \filipolong\ system which automatically finds connections (i.e., linkage points) between data provided by APIs and local knowledge bases. \filipo\ is an open source sample-driven schema matching system that models API services as parameterized queries. Furthermore, our approach is able to find valid input values for APIs automatically (e.g. IDs) and can determine valid alignments between KBs and APIs. Our results on ten pairs of KBs and APIs show that \filipo\ performs well in terms of precision and recall and outperforms the current state-of-the-art system.
\end{abstract}

\section{Introduction}\label{sec:introduction}
RDF knowledge bases (KBs) are used in many domains, e.g bibliographic, medical, and biological data. Most knowledge bases face the problem that they are potentially incomplete, incorrect or outdated. Considering how much new data is generated daily it is highly desirable to integrate missing data provided by external sources. Thus, data integration approaches~\cite{koutraki_cikm_2015,koutraki_eswc_2017,qian_2012,bernstein_2011,madhavan_2001} are used to expand KBs and correct erroneous data. The usual process of data integration is to download data dumps and align the schemas of a local KB and these dumps. ``Aligning" describes the process by which relations and classes from the local KB are mapped to relations and entities of external sources, thus creating a mapping between the local and the external data schemas. Afterwards, the integration process can be done and the data of the KB is expanded or updated.

However, data dumps are often updated only infrequently. Using live data through APIs instead of dumps~\cite{koutraki_cikm_2015,koutraki_eswc_2017} allows access to more recent data. In addition, the number of potential data sources becomes much larger when using APIs since most data providers share their data not via dumps, but via APIs. According to Koutraki et al.~\cite{koutraki_cikm_2015}, \emph{APIs seem to be a sweet-spot between making data openly accessible and protecting it}. The problems of data integration, i.e. how two different schemas can be mapped, remain. In the worst case, the schema of an external source has a completely different structure than the local KB. Hence, data integration remained a manual task for most parts~\cite{koutraki_cikm_2015}.

\textbf{Motivation.} Connecting KBs with data behind APIs can significantly improve existing intelligent applications. As a motivating example, we consider dblp\footnote{\url{https://dblp.uni-trier.de/}}, a bibliographic database of computer science publications. It accommodates different meta data about publications, e.g., titles, publisher names, and author names and can be represented as an RDF KB. Data from dblp is often used for reviewer, venue or paper recommendation, and extending dblp with information from APIs like CrossRef, or SciGraph, for example titles or abstracts, can improve these applications. Missing information about authors like ORCIDs (an ORCID is a code to uniquely identify scientific authors) can be supplemented by these APIs and help to disambiguate author profiles. Furthermore, such information is also useful for a user querying dblp for authors or publications, where missing information can be completed using external data sources. Therefore it is important that multiple APIs can be used and missing data can be integrated from many different sources. Additionally, the determined alignments can be used to identify erroneous data and correct it if necessary.

\textbf{Contributions.} We present \filipolong~\footnote{Code available at \url{https://github.com/dbis-trier-university/FiLiPo}}, a system to automatically discover alignments between KBs and APIs, focusing on detecting property/path alignments. We omit aligning classes because classes and types (in terms of semantic classes, e.g. URLs) do not exist in typical API responses. \filipo\ is designed to work with single record response APIs, i.e. APIs that return only a single record as response and not a list of most similar search results, and works for datasets of arbitrary domains. In contrast to other systems~\cite{qian_2012}, users of \filipo\ only require knowledge about a local KB (e.g. class names) but no prior knowledge about the APIs' data structure. To the best of our knowledge, \filipo\ is the first aligning system that automatically detects what information from a KB has to be used as input of an API to retrieve responses. Thus end users do not have to determine the best input, significantly reducing manual effort. In contrast to other state-of-the-art systems~\cite{koutraki_cikm_2015}, \filipo\ uses fifteen different string similarity metrics to find an alignment between the schema of a KB and that of an API. A single string similarity method is not suited to compare different kinds of data, for example both ORCIDs (requiring exact matches), ISBNs (with some variation) and abbreviated names. A user only needs to specify the number of requests sent to the API in order to keep the approach simple.

\section{Problem Statement}\label{sec:problem_statement}
This paper addresses five challenges when aligning local KBs with APIs. The first challenge is to determine which input values (e.g. DOI, etc.) have to be sent to the API to retrieve a valid response. A valid response is a response that contains information about the requested entity. In contrast, invalid responses contain information about similar entities (e.g. a list of most similar search results) or an error message. Note that the user has to specify the URL and the parameter of an API (e.g. \texttt{www.example.com/api?q=}). When a resource is requested that is unknown to the API, it can respond in several ways. The classic case is that it returns an HTTP status code (e.g. \texttt{404}). The more complicated case is a JSON response that contains an error message or returns information on a ``similar" resource (e.g., with a similar DOI). This cannot be easily distinguished from a ``real" response which contains data about the requested resource. 

An alignment between the schemata of an API and a KB are determined by collecting several responses from an API and comparing these information with the one stored in a KB. Semantically equal data values between API responses and KB entities are denoted as (sample) matches (e.g. the match of a DOI value). In order to determine such matches, the second challenge is that the same value may be represented slightly differently in the KB than in the API (e.g., names with and without abbreviated first names), hence the comparison needs to apply string similarity methods. The various existing similarity methods have different strengths and weaknesses. For example, Levenshtein distance is good for comparing the titles of a paper or movie, but performs poorly when comparing names of authors or actors because names are often abbreviated and first and last names may be in different order. Hence, a suitable similarity method needs to be determined automatically for each type of data.

A special case of this challenge is comparing identifiers, e.g. ISBNs. Identifiers need to be equal in order to yield as match. However, the ISBN of a book can be written in different forms (e.g. without hyphens) but should be considered equal. For this reason a simple check for equality is not sufficient, otherwise possible alignments are lost. Note that such identifiers (e.g. IBANs, tax numbers and others) also exist in other domains.

Finding a match between the information of KBs and APIs can be particularly problematic if APIs respond with records similar to the requested one. For example, a request for a book with title ``\emph{Some example Title}" may lead to an API response of a book with title ``\emph{Some Title}". The information of the API and the KB may overlap, especially for values that appear in many entities (e.g. year). Thus, the fourth challenge is to check if APIs respond with the requested information. Koutraki et. al~\cite{koutraki_cikm_2015} state that if KBs and APIs share the same domain, it is likely that the data of their entities overlap. This means that if the information of the API and the KB overlaps sufficiently, the API has probably responded with the requested record.

The last challenge is that some data values are contained in an API response several times, e.g. year values. In this case, they may represent a different piece of information, e.g. some bibliographic APIs respond with data containing references and citations of a paper, which often include author names and publication years. During the matching process, care must be taken as to which information is matched. Just because the values match, they do not form a valid match (e.g. matching a papers author names with the author names of the papers references). Hence the semantics and structure of the paths should be considered but API responses do not always have a clear or a directly resulting semantics.

\section{Related Work}\label{sec:related_work}
\textbf{Web API Alignment.} Some work has already been done regarding the aligning of different datasets. However, only the DORIS system~\cite{koutraki_cikm_2015,koutraki_iswc_2015} has dealt with the alignment of KBs and APIs so far and is used as baseline system for the evaluation of \filipo. The core idea of DORIS is to build upon the schema and structure of an existing KB containing instances and facts about these instances. First, the system sends some probing requests to a chosen API. It uses label information of instances as its predefined input relation for APIs. However, this is not always the appropriate input parameter for the API. For example, some APIs expect DOIs or ISBNs as input parameters.

One key assumption of the DORIS system is that it is more likely to find information on well-known, popular or famous entities (e.g. famous actors, acclaimed books, or big cities) via APIs calls than it would be for unknown entities. Additionally, Koutraki et al. assume that KBs contain more facts (triples) for well-known entities and therefore rank the entities of the input class by descending number of available facts. While this is a reasonable approach for open-topic knowledge bases like YAGO, it is likely to fail for domain-specific KBs. For a publication, for example, the number of facts stored by a bibliographic KB is often determined by the meta data of that publication, not by its popularity (unless citations etc. are stored). In contrast to this approach \filipo\ picks randomly chosen entities of a KB. This is done to prevent the entities from being very similar to each other and thus increase the probability of an response. For example, assuming that an API only responds to entities with a specific publisher, e.g. Springer. If entities are selected in any non-random way, e.g. according to the amount of facts like in the DORIS system, it is possible that no entities with publisher Springer are included, and the API cannot respond. As a consequence no aligning can be done and the approach will be unsuccessful.

DORIS normalises values by lower casing strings, ignoring punctuation, and reordering the words in alphabetical order. A benefit of this approach is that a pairwise comparison of all combinations of KB and API values is no longer necessary. A relation and a path are considered as a match if the corresponding values are exactly equal after normalisation. Therefore, the system can use an efficient merge-join algorithm. Normalised values of the paths and relations are sorted in alphabetic order. Afterwards, the two lists can be merged to produce the relation-path matches. This stands in contrast to our approach since, as stated previously, a single method is not suited to compare all various data types. The limitations of DORIS become clear when examining, for example, author names or titles. Names are often abbreviated and the matching approach will fail because DORIS performs an exact match on the normalised names. Similar problem will arise when examining other data values containing typos or minor differences. In contrast to this approach \filipo\ uses a set of similarity methods and picks randomly chosen entities of a KB.

\textit{Wrapper Inference.} The problem of aligning KBs and APIs shares similarities with various other fields~\cite{rahm_survey_2001,bernstein_2011,koutraki_cikm_2015} like schema alignment, query discovery and wrapper inference approaches. Wrapper inference approaches~\cite{senellart_widm_2008,derouiche_icde_2012} face similar problems as alignment systems from other fields. Senellart et al.~\cite{senellart_widm_2008} present an approach which uses domain knowledge (concept names and instance data) in order to identify the input of form fields. They assume that there are no specifically required fields in the form. However, this does not apply to the majority of APIs, since most have a mandatory parameter. Afterwards the structure of the data behind the form fields is aligned with concept names by exploiting the semantics of form fields and web tables (e.g. labels, table headers, etc.). Since paths in API responses do not always have a clear or any semantic at all, \filipo\ does not use path semantics. Derouiche et al.~\cite{derouiche_icde_2012} also use domain knowledge to extract data from Web sources. Additionally, they use for every concept (e.g. date) a form of regular expression. Since users have to specify these expressions, this approach significantly raises the manual effort and the needed knowledge.

\textbf{Schema/Ontology Alignment.} Aligning data of KBs and APIs has similar problems as schema/ontology alignment. The major difference is that API responses often do not have explicit semantics or any semantics at all, and the data schema of the API is often not directly accessible to external parties. In addition, names of the paths are often ambiguous. Semantics in form of rules (as with RDF/OWL) does not exist in API responses. Also, responses usually do not provide information about classes/relations that can be used for the alignment process. When using APIs, only instance information is available and hence classical schema/ontology approaches are not suitable. Additionally, Madhavan et al.~\cite{madhavan_2005} state that KBs often contain multiple schemas to materialise similar concepts and hence build variations in entities and their relations. This makes schema-based matching inaccurate, which must therefore be supported by evidence in form of instances. 

\textit{Machine Learning.} Machine Learning is experiencing a big boom in the past years. Approaches like BERT\cite{bert2018} or Flair\cite{flair} (both frameworks that learn contextual relations between words in a text) enable to grasp the semantics of data and to compare them with others. A clear disadvantage of machine learning alignment approaches~\cite{schmidts_2019,sahay_2020,koutraki_eswc_2017} is that they require training data, which is either provided by experts or learned from predefined mappings. Even though the results look promising, the approach has the problem that sufficient training data must be available. However, there is only a small number of available alignments and data sources. In addition, the training data must cover many different data structures and different hierarchical data types from many different domains.  Therefore, this approach is currently still subject to strong limitations. In contrast, DORIS and \filipo\ do not require the user to be a machine learning expert and assume no assistance through experts during the mapping process. Additionally, the results are often worse than with a rule-based program or classic approaches. 

\textit{Schema Alignment.} Classical approaches for schema or ontology alignment take a set of independently developed schemas as input and construct a global view~\cite{aumueller_2005,cruz_2009} over the data. Since the schemas are independently developed, they often have different structure and semantics. This is especially the case when the schemes are from different domains, of different granularity or simply developed by different people in different real-world contexts~\cite{rahm_survey_2001}. The first step when performing schema alignment is to identify and characterise relationships that are shared by the schemas. After identifying the shared relations they can be included in a global schema. Since APIs do not provide any schema information it is not possible to use classical schema or ontology alignment strategies. 

Approaches like COMA++\cite{aumueller_2005} represent customisable generic matching tools for schemas and ontologies. COMA++ relies on a taxonomy to determine an alignment and uses a so-called fragment based match approach (breaking down a large matching problem into smaller problems). Additionally, it enables various interaction possibilities to the user in order to influence the alignment process. Furthermore, COMA++ does not use any instance information of the KBs. Madhavan et al.~\cite{madhavan_2005} state that KBs often contain multiple schemas and data models to materialise similar concepts and hence build variations in entities and their relations. This makes purely schema-based matching inaccurate, which must therefore be supported by evidence in form of instances from the KB. \filipo\ accomplishes this by working with instances from the KB and comparing the actual values, not just determining a match at the schema level.

Systems like BLOOMS~\cite{jain_2010} use schema information if it is present in the Linked Open Data (LOD) cloud. This means that a schema no longer needs to be explicitly declared and the schema information of reused vocabularies (ontologies) can be accessed via the LOD cloud. However, since API responses (mostly) do not contain explicit schema information (of the LOD domain), the procedure is not applicable. In addition, the collected answers of an API implicitly contain a schema which only needs to be extracted~\cite{benedetti_2014}. This is similar to KBs which contain implicit schema information even without the specification of a schema. To keep the process as simple as possible much user assistance and knowledge should not be required. Furthermore, our goal is a fully automatic determination of an alignment without any user assistance during the process.

Systems like AgreementMaker~\cite{cruz_2009} are based on the assumption that users of the system are domain experts and thus build on user assistance. The system uses a large set of matching methods and enables the user to define the types of components to be used in the aligning process, i.e. using the schema only or both schemas and instances. 

\textit{Instance-Based Alignment.} Instance-based alignment systems use the information bound to instances in KBs in order to find shared relations and instances between two KBs. These approaches can be divided into instance-based class alignment approaches and instance-based relation alignment approaches. The main difference between class and relation alignment lies in the fact that relations have a domain and range. Even if relations share the same value, they can have different semantics (e.g. \texttt{editor} and \texttt{author}).

A lot of works~\cite{qian_2012,dhamankar_2004,madhavan_2005,koutraki_edbt_2016} focus on instance-based relation alignment between two KBs. However, most of them focus on finding 1:1 matches, e.g. matching \texttt{publicationYear} to \texttt{year}. The iMAP system~\cite{dhamankar_2004} semi-automatically determines 1:1 matches, but also considers the complex case of 1:n matches. iMAP consists of a set of search modules, called searchers. Each of the searchers handles specific types of attribute combinations (e.g. a text searcher). \filipo\ follows a similar approach. Additionally, iMAP offers users an explanation of the determined match containing information about why a particular match is or is not created, and why a certain match is ranked higher than another. Instead of searchers, \filipo\ only distinguishes between the type of information (numeric, string, or is it a key). Then, in case of strings, a number of different similarity methods are used, and the best method is automatically determined and used. 

Similar to iMAP, MW\textsc{eaver}~\cite{qian_2012} also needs user assistance. MW\textsc{eaver} realises a sample-driven schema mapping approach which automatically constructs schema mappings from sample target instances given by the user. The idea of this system is to allow the user to implicitly specify mappings by providing sample data. However, this approach needs significant manual effort. The user must be familiar with the target schema in order to provide samples. In contrast to this approach, \filipo\ draws the sample data randomly from the KB and thus tries to cover a wide range of information.

SOFYA~\cite{koutraki_edbt_2016} is an instance-based on-the-fly approach for relation alignment between two KBs. The approach works with data samples from both KBs in order to identify matching relations. The core aspect of SOFYA is that the standard relation ``sameAs" is used to find identical entities in two different KBs. However, this mechanism cannot be used for the alignment of KBs and APIs, because APIs do not contain standardised ``sameAs'' links. 

\textit{Holistic Approaches.} Some developed systems cannot be categorised easily since they use techniques of all fields. ILIADS~\cite{udrea_2007} for example takes two OWL ontologies as input and determines afterwards an alignment based on lexical, structural and extensional similarity. They combine a flexible similarity matching algorithm with an incremental logical inference algorithm. Additionally, ILIADS uses a clustering algorithm which considers relationships among sets of equivalent entities, rather than individual pairs of entities. 

\textsc{PARIS}~\cite{suchanek_2011} is an instance-based approach that aligns related entities and relationship instances, but also related classes and relations. The goal of \textsc{PARIS} is to discover and link identical entities automatically across ontologies. It was designed such that no training data or parameter tuning is needed. \textsc{PARIS} does not use any kind of heuristics on relation names and therefore it is able to align relations with completely different names. However, a downside of \textsc{PARIS} is that it is not able to deal with structural heterogeneity. This is a major problem, because most KBs are structured differently from each other, since they have been developed by different people~\cite{rahm_survey_2001}.

Madhaven et. al. developed a system called Cupid~\cite{madhavan_2001} which is used to discover an alignment between KBs based on the names of the schema elements, data types, constraints and structure. It combines a broad set of techniques of various categories (e.g. instance-based, schema alignment, etc.). The system uses an linguistic and structural approach in order to find a valid alignment. 

\section{Preliminaries}\label{sec:preliminaries}
\begin{figure}[t]
\centering
\subfloat[Response Example.\label{fig:request_response}]{
  \includegraphics[scale=0.55]{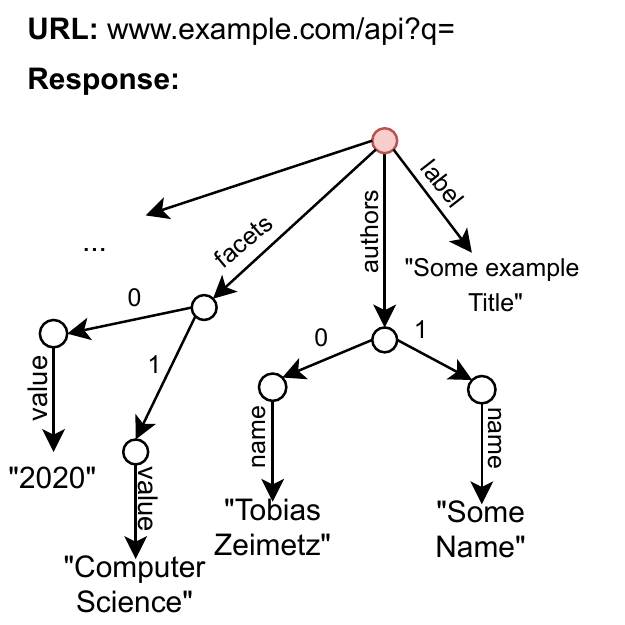}
}\hfil
\subfloat[Fragment of an RDF KB.\label{fig:knowledge_graph}]{%
  \includegraphics[scale=0.45]{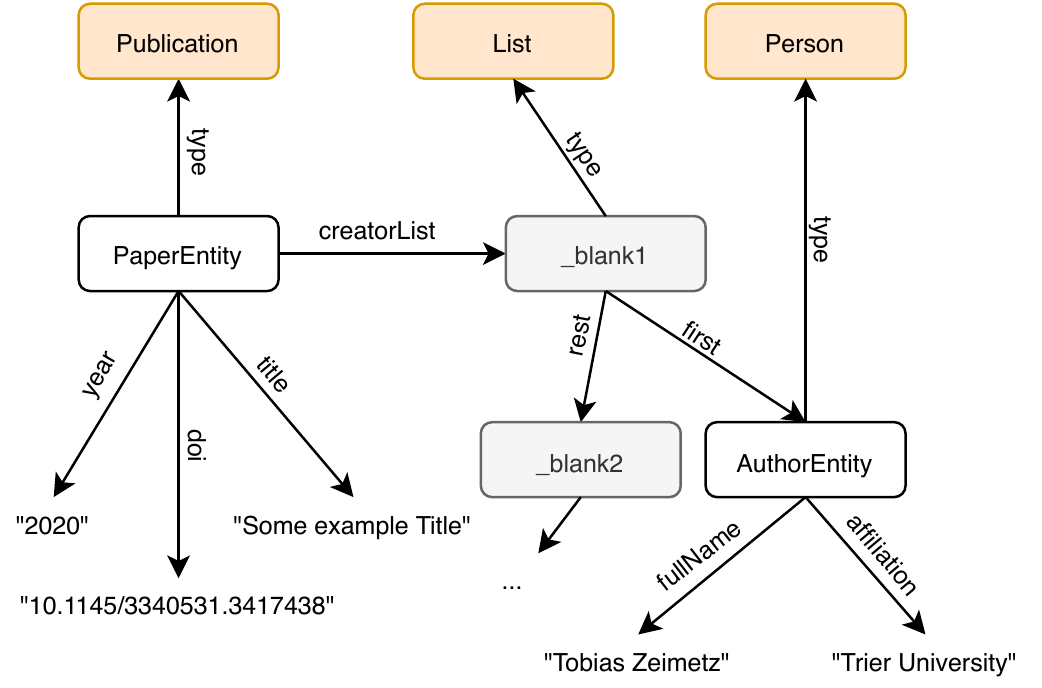}
}

\caption{Record of a KB and the corresponding API response.}
\label{fig:records}
\end{figure}

\textbf{Knowledge Bases.} An RDF KB can be represented as graph with labelled nodes and edges. A KB consists of triples of the form $(s,p,o)$ which represent a fact in the KB. The subject (start node) $s$ describes the entity the fact is about, e.g. a paper entity. An entity is represented by an URI, which is an identifier of a real-world entity such as a paper or an abstract concept (e.g. a conference). The predicate $p$ describes the relation between the subject and the object, e.g. \texttt{title}. The object (target node) $o$ describes an entity, e.g. an author, or is a literal, e.g. the title of a publication. A class in a KB is an entity that represents a group of entities. Entities assigned to a class are instances of that class. In Figure~\ref{fig:knowledge_graph}, the entity \texttt{PaperEntity} is an instance of the class \texttt{Publication}.

\textbf{Relations.} Since this paper focuses on aligning relations, we introduce a formal definition for relation (paths). Given a KB $K$, if $(s,r,o)\in K$, we say that $s$ and $o$ are in relation $r$, or formally $r(s,o)$; in other words, there is a path from $s$ to $o$ with label $r$. Also, we write $r_1.r_2.....r_n(s,o)$ to denote that there exists a path of relations $r_1,r_2,...,r_n$ in $K$ from subject $s$ to object $o$ visiting every node only once. For example, in Figure~\ref{fig:knowledge_graph} the relation \texttt{year(PaperEntity,"2020")} describes the path from the entity \texttt{PaperEntity} to the value \texttt{"2020"}. In the following we will refer to $r_1.r_2.....r_n(s,o)$ as relation-value path.

\textbf{Identifier Relations.} Some KBs contain globally standardised identifiers such as DOIs, IBANs (International Bank Account Numbers), tax numbers and others. Identifiers are only bound to a single entity and should be unique. Therefore, relations $r$ that model identifier relations have the constraint that their inverse relations ($r^{-1}$) are ``quasi-functions", i.e., their inverse relations have a high functionality. Many works~\cite{koutraki_cikm_2015,suchanek_2011,hogan_nefors_2010} have used the following definition for determining the functionality of relations:
\begin{equation*}
    fun(r) \coloneqq \frac{|\{x : \exists y : r(x,y)\}|}{|\{(x,y) : r(x,y)\}}
\end{equation*}

Since real world KBs are designed by humans identifier relations are often error-prone which is why some identifier values may appear more than once. Hence, we consider every relation $r$ contained in $K$ with $fun(r^{-1}) \geq \theta_{id}$, where $\theta_{id} \in [0,1]$ is a threshold, as \textit{identifier relation}. From now on we denote $R^{id}_{K}$ as the set of all identifier relations (e.g., \texttt{doi}, \texttt{isbn}, etc.) contained in a KB $K$ . Note that we ignore identifier relations that are composed of several relations.

\textbf{Web API.} A Web service can provide one or multiple APIs to access data. APIs are called via parameterised URLs (see Figure~\ref{fig:request_response}). As shown in Figure~\ref{fig:request_response} the response of an API can be represented as an unordered and labelled tree. Inner nodes in the tree represent an object (similar to an entity in a KB) or an array, leaf nodes represent values. The path to a node represents a relation between an instance (similar to an entity in a KB) and another instance or value. To avoid confusion we will describe these relations in a response only as paths. 

\textbf{Path-Value-Pairs.} In order to find valid alignments between KBs and APIs the information in the API responses has to be compared with the values of the corresponding entities in a KB. Since comparing objects and arrays from the API response with entities from the KB to determine alignments is not promising, only paths to leafs (literals) have to be considered. Given an API response $res$ we will write $p_1.p_2.....p_n(o)$ to denote that there is a path $p_1,p_2,...,p_n$ in $res$ from the root of the response to the leaf $o$ with these labels. For example, in Figure~\ref{fig:request_response} the path \texttt{label("Some example Title")} describes the path from the root of the API response to the leaf \texttt{"Some example Title"} via the path \texttt{label}. 

\textbf{Branching Points.} A branching point in an API response indicates that there are several outgoing edges from one node, labelled by numeric index values 0 to n. These branching points represent arrays. In the example in Figure~\ref{fig:request_response}, the path \texttt{authors.0.name("Tobias Zeimetz")} contains a branching point. To indicate a branching point, we will use the symbol \texttt{*} instead of the numeric index in paths; in the example, we will write \texttt{authors.*.name("Tobias Zeimetz")}. Using the same logic, a relation in a KB that points to a set of entities is considered to be a branching point (e.g. \texttt{creatorList} in Figure~\ref{fig:knowledge_graph}). Additionally, we write $P*$ to indicate a path $P*p$ that has $P$ as prefix and $p$ as suffix, with a branching point separating the two parts.

\section{System Overview}\label{sec:system_overview}
\begin{figure}
    \centering
    \includegraphics[scale=0.7]{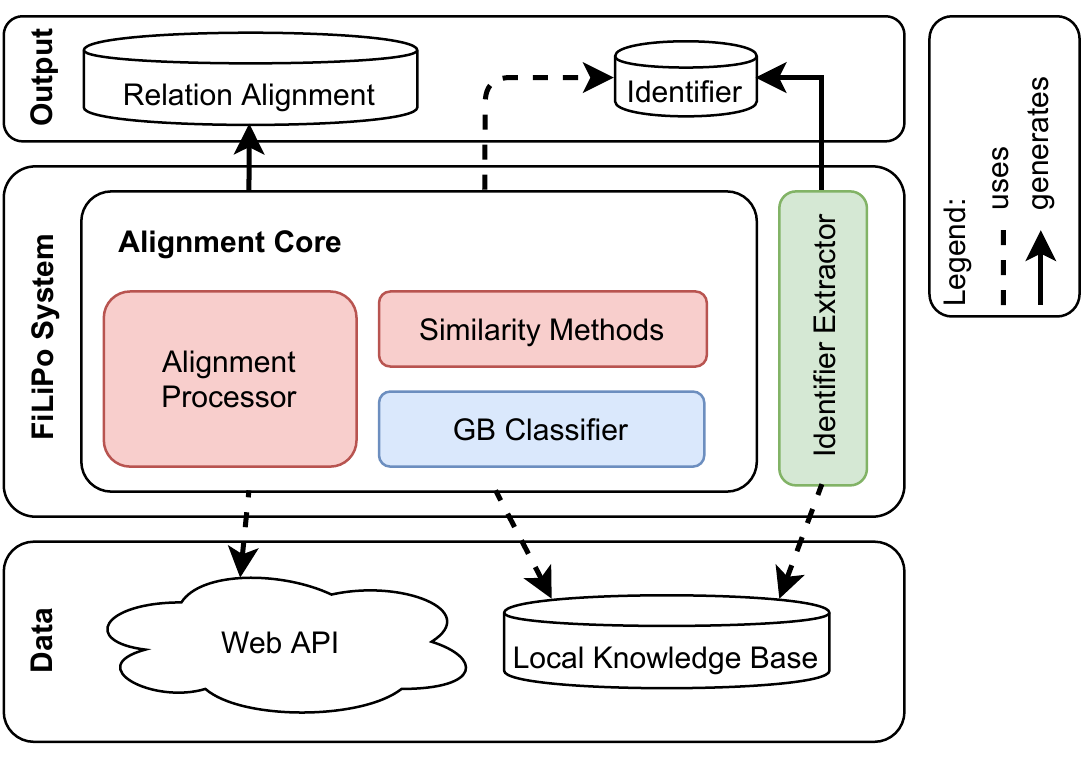}
    \caption{\filipo\ System Architecture.}
    \label{fig:system_overview}
\end{figure}

In order to better understand how \filipo\ proceeds, we will first give an overview of the system components. The \filipo\ architecture shown in Figure~\ref{fig:system_overview} is divided into two main components, the \textit{Alignment Core} and the \textit{Identifier Extractor}. Additionally, the system uses a set of configuration files to manage local KBs and APIs that will be used for the alignment process. 

\begin{figure}[t]
\centering
\subfloat[KB Management.\label{fig:kb_management}]{
  \includegraphics[scale=0.42]{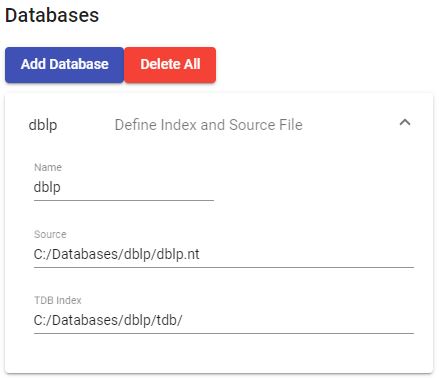}
}\hfil
\subfloat[API Management.\label{fig:api_management}]{%
\includegraphics[scale=0.47]{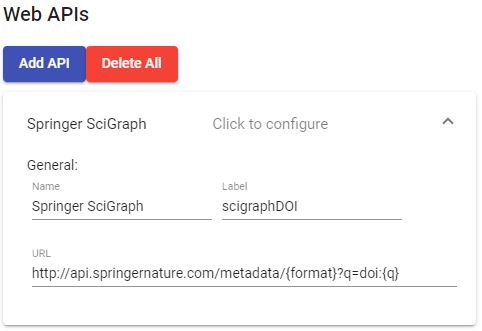}
}
\caption{Example Screenshots of FiLiPos Data Management.}
\label{fig:data_management}
\end{figure}

\textbf{Configuration Files.} There are two types of settings in \filipo. The first one (\texttt{Global Settings}) stores all global settings like the paths to output files and a set of different thresholds values (more in in Section~\ref{sec:matching_mapping}). The second setting (\texttt{Data Management}) stores information about KBs and APIs. It is used to manage multiple data sources and therefore stores the paths to registered KBs and the URLs (see Figure~\ref{fig:data_management}) under which the registered APIs can be accessed. Additionally, the input class (e.g. \texttt{Publication}) for every API that is used is specified. The input class describes the class of entities from the used KB whose facts are used to request information from the API. For example, the API presented in Figure~\ref{fig:api_management} has as input class \texttt{Publication} and values of the input relation \texttt{doi} are used to request the API.

\textbf{Identifier Extractor.} The Identifier Extractor is used to derive the identifier relations (e.g. ISBNs, ISSNs, etc.) contained in a given KB. Therefore, the system computes the functionality of each relation. Since real world KBs may be noisy and contain errors, a perfect functional relation is unlikely. Therefore, we determined the functionality of all relations in a KB. In case a relation has a functionality greater or equal to 0.99 (in case the KB contains erroneous data), the system assumes that the relation describes an identifier. It is important to determine the identifiers of the KB because the \filipo\ system uses several similarity methods that may be too fuzzy to compare identifier values. However, this explained in more detail in Section~\ref{sec:aligning}.

\textbf{Alignment Core.} The \textit{Alignment Core} consists of three components; the \texttt{Alignment Processor}, the \texttt{Similarity Processor} and the gradient boosting classifier (\texttt{GB Classifier}). The Alignment Processor interacts with the other components to determine a correct alignment. The \texttt{Similarity Processor} is used by the \texttt{Alignment Processor} to compare values of a KB record and an API response. \filipo\ uses the string similarity library developed by Baltes et al.~\cite{sotorrent_msr_2018} in order to compare these values and to find valid matches. This similarity library uses three types of similarity methods: (1) equal, (2) edit and (3) set-based. String similarity methods of the equal category check for the equality of two strings; edit-based methods (e.g. Levenshtein) define the similarity of two strings based on the number of edit operations needed to transform one string into the other~\cite{sotorrent_msr_2018}. Set-based methods determine how large the intersection of two strings in terms of tokens (e.g. n-grams, n-shingles, etc.). We excluded the overlap method in the set-based category since this method is too fuzzy and would lead to an erroneous aligning (e.g. aligning a title and an abstract since they have in most cases many tokens in common). By using this library \filipo\ can use up to 15 different similarity methods with several variants. 

\textbf{Gradient Boosting Classifier.} In addition to the similarity methods, the \texttt{Alignment Processor} uses a specialised method to compare identifier values (\texttt{GB Classifier}). For example, in order to match an ISBN with a value returned by an API the values need to be equal. As explained previously, identifiers can have several writing styles and therefore need to be normalised before comparing, e.g. removing hyphens in ISBNs. Using fuzzy similarity methods like Levenshtein is inaccurate but using \textit{equals} is too strict. Therefore \filipo\ utilises a gradient boosting classifier working on Flair~\cite{flair} embeddings of identifiers to determine whether two (identifier) values are equal. We use Flair embeddings instead of other embeddings since this framework is character-based and therefore suits better for the comparison of two identifier strings. A detailed explanation of the procedure of the Classifier can be found in Section~\ref{sec:aligning}.

\section{Schema Matching and Mapping}\label{sec:matching_mapping}
\filipo\ operates in two phases. First (\emph{probing phase}), \filipo\ sends various information (e.g. DOIs, titles, etc.) to an API to determine which information the API responds to. Afterwards (\emph{aligning phase}), the information returned is used to guess the APIs schema and to determine an alignment between the local and external data. The input of the aligning process is the URL of the API (see Figure~\ref{fig:request_response}) and the corresponding input classes in the KB. An input class is a class of entities that will be used to request the API.

\begin{figure}[t]
    \centering
    \includegraphics[scale=0.8]{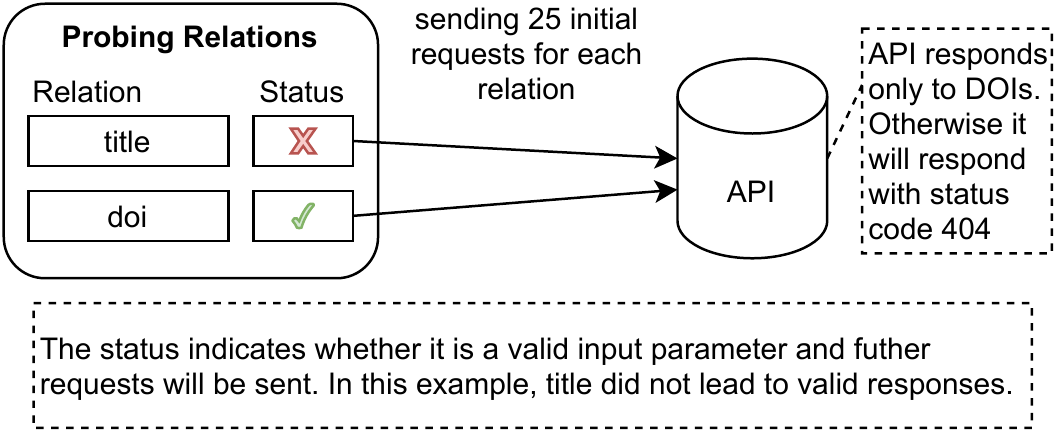}
    \caption{Short Example of the Probing Process.}
    \label{fig:probing}
\end{figure}

\subsection{Probing Phase}
The probing phase is used to find the set $R_{in}$ of relations of the input class that point to values which can be used to request the API successfully (e.g., a DOI relation). To illustrate this with an example (see Figure~\ref{fig:probing}), we assume that the input class of the API whose result is presented in Figure~\ref{fig:request_response} is \texttt{Publication}. The illustrated fragment in Figure~\ref{fig:knowledge_graph} has five relations to describe the metadata of a publication but the API only responds to DOIs. First, all relations that are not connected to literals (e.g., \texttt{type}) are ignored. In addition, values that occur more than once in the KB (year numbers, ambiguous titles and names) are not used as input values. The reason for this is that more than one record can be returned and possibly a not matching one is returned. Afterwards $n_p$ initial requests are sent to the API for each remaining relation of the input class (i.e. \texttt{title} and \texttt{doi}). The created URL to request the API is a concatenation of the API URL specified by the user (see Figure~\ref{fig:request_response} and values for the corresponding input relations, e.g. for \texttt{doi} an example call URL is \texttt{www.example.com/api?q=10.1145/3340531.3417438}.

The input values for each relation are picked uniformly at random from entities of the input class in the KB. As explained previously, this is done to prevent the entities from being very similar to each other and thus increase the probability of an response. For example, assuming that an API only responds to entities with a specific publisher, e.g. Springer. If entities are selected in any non-random way, e.g. according to the amount of facts, it is possible that no entities with publisher Springer are included, and the API cannot respond and no aligning can be done.

\begin{figure}[t]
    \centering
    \includegraphics[scale=0.5]{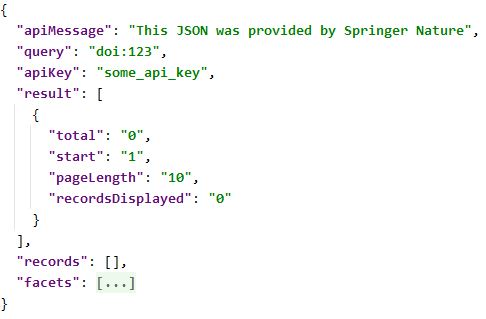}
    \caption{Example of an API Error Message.}
    \label{fig:error_message}
\end{figure}

After sending a request to the API it can respond in several ways. In the best case, the API responds with the HTTP status code \texttt{200 OK} or with an HTTP error code (e.g. \texttt{404 Not Found}). In the worst case, the server responds with a document containing an error message (see Figure~\ref{fig:error_message}). In this case the system cannot easily detect that some input values did not lead to a (successful) response and therefore will continue with the alignment phase with the corresponding relation. Since this would result in a considerable increase of requests and runtime it is important to identify error messages.

In order to identify error responses, the system iterates over all answers and compares how similar they are to one another. This procedure is based on the observation that error responses are always similar or even the same, i.e. they usually contain the same error message or consist of an generic error message in combination with the request value. An example of this assumption is shown by the error response of SciGraph in Figure~\ref{fig:error_message}. The error response of SciGraph always looks similar. The only difference is in the field ``query", because it will always contain the value which was used to request SciGraph. In contrast, correct answers are different to one another since they contain information about various different entities. As a result, an error response is determined by counting how often a response was similar (by using Levenshtein) to other responses. The one that is most often similar (i.e. the similarity is higher than 0.80) to other responses is considered an error message. Then all responses similar to this response will be deleted and all relations $r_{in}$ which have not received enough answers will no longer be considered as valid input relations. In this way unnecessary requests are prevented. 

Next, the alignment phase begins, considering only the set $R_{in}$ of relations that led to valid answers (for the example in Figure~\ref{fig:probing} only relations with a green check mark will be used in the next phase). The aligning phase itself is divided into two parts: (1) determining candidate alignments and (2) determining the final alignments.

\subsection{Aligning Phase: Candidate Alignment}\label{sec:aligning}
This phase takes as input the set of valid input relations $R_{in}$, a KB $K$ and the corresponding identifier relations $R^{id}_{K}$.

\begin{figure}[t]
\centering
\includegraphics[scale=0.6]{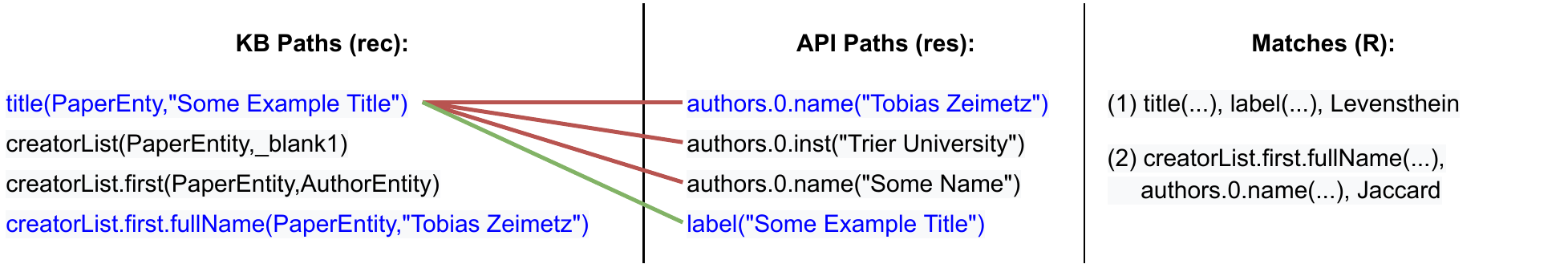}
\caption{Fragment of a KB Record and an API Response.}
\label{fig:paths}
\end{figure}

For each input relation $r_{in}\in R_{in}$, the algorithm sends $n_r$ further requests to the API. A random entity $e$ is chosen from the input class. \filipo\ then retrieves the set $rec$ of all facts that $K$ contains for $e$ in form of relation-value paths $r(e,l$) (see Figure~\ref{fig:paths}, left side). Note that $r$ can be a path of multiple relations, e.g. $r_1.r_2...r_n(e,l)$. Like Koutraki et al.~\cite{koutraki_cikm_2015} we take only facts into account up to depth three. This depth was chosen because all other facts usually do not make statements about the entity $e$. To exclude the case that entities are connected to other entities in only one direction, inverse relations are also considered. Afterwards \filipo\ calls the API with values $v_{req}$ of the input relation $r_{in}$ of $e$ and stores the response in $res$ (see Figure~\ref{fig:paths}, middle part). Note, that for the sake of simplicity not all KB and API paths are shown in Figure~\ref{fig:paths}.

The next step is to find all relation matches $R$ between $rec$ and $res$ (see Figure~\ref{fig:paths}, right side). The set $res$ encodes information from the response as path-value pairs of the form $p(v)$ where $p$ is the path in the response from the root to the value $v$. Note that $p$ can be a path of multiple components, e.g. $p_1.p_1...p_n(v)$. All values $l$ of $rec$ must be compared with all values $v$ of $res$. Figure~\ref{fig:paths} presents an example for the title relation. The coloured lines represent comparisons between the values, red lines denote invalid matches and the green line represents a valid match. For each pair ($r(e,l),p(v)$), a suitable similarity method is determined. If $l$ or $v$ is an IRI, it is important that they are compared with \texttt{equals} as IRIs are identifiers and hence only the same if they are identical. The same holds for numerical values. In all other cases \filipo\ uses a set $M_{sim}$ of fifteen different similarity methods\footnote{All used similarity methods are listed in our manual at \url{https://github.com/dbis-trier-university/FiLiPo/blob/master/README.md}} with several variants since one string similarity method is not sufficient to compare several data types. The method $m \in M_{sim}$ returning the largest similarity of $l$ and $v$ is considered (temporarily) to be a suitable method to compare both values and is stored for the later process. 

We used the string similarity library developed by Baltes et al.~\cite{sotorrent_msr_2018} since it contains all major similarity methods, divided into three categories: equal, edit and set based. Since fuzzy methods are not appropriate for identifier relations and comparing them for equality would be too strict, identifier relations in $R^{id}_{K}$ are therefore compared with a gradient boosting classifier working on Flair~\cite{flair} embeddings. We use Flair embeddings since this framework is character-based and therefore suits better for the comparison of identifier values. Once the best similarity method has been determined, and if it yields a similarity of at least $\theta_{str}$, the triple $(r,p,m)$ is created and added to the set of record matches $R$.

\begin{figure}[t]
\centering
\includegraphics[scale=0.8]{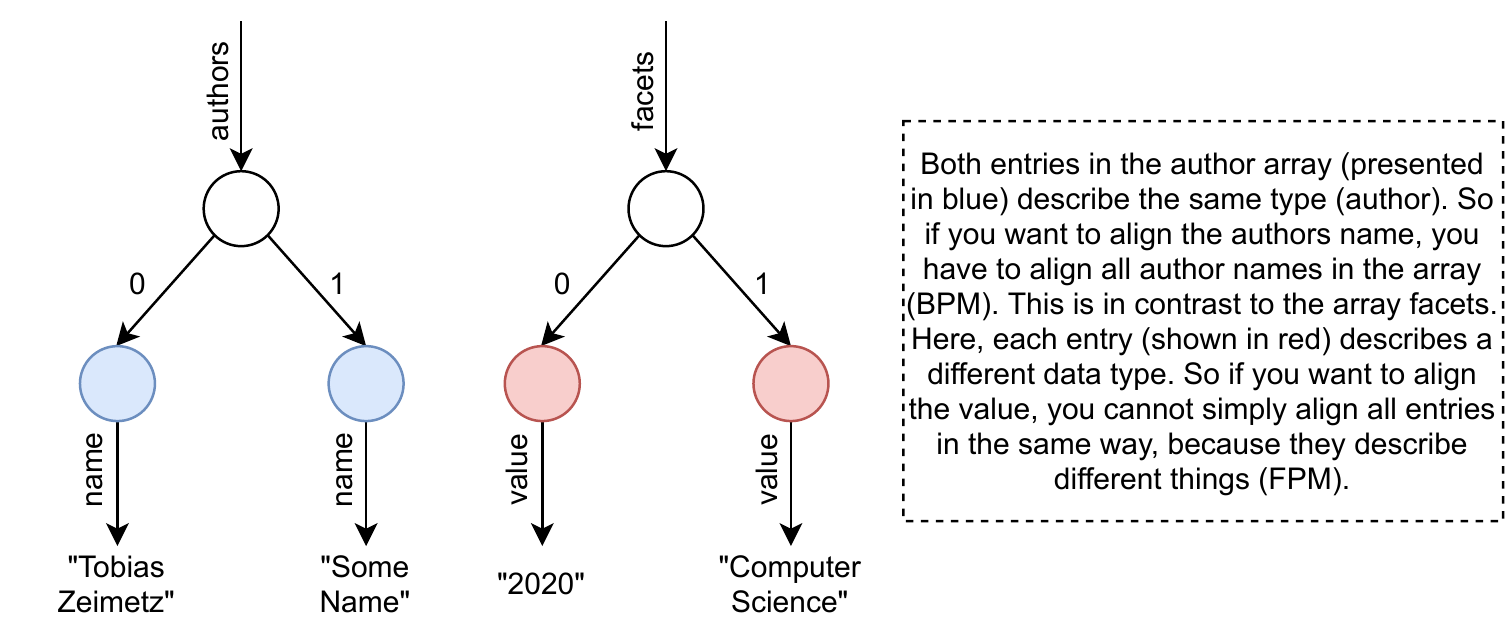}
\caption{Explanation of BPM and FPM.}
\label{fig:branching_example}
\end{figure}

If enough matches are found, it is assumed that the input entity $e$ and the API response overlap in their information (the overlapping information is highlighted in Figure~\ref{fig:paths} in blue) and that the API has responded with information about the requested entity. We compute the overlap by dividing the number of matches $|R|$ by the number of entries of the smallest record $rec$ or $res$.  If the overlap is greater than a threshold $\theta_{rec}$, the overlap is considered sufficient and the matches $R$ will be added to $A_{r_{in}}$ (an example of overlapping values/paths is presented in blue in Figure~\ref{fig:paths} and an example of the set  $A_{r_{in}}$ is given in Figure~\ref{fig:alignment}). This set represents matches found for the input relation $r_{in}$. If not enough matches are found, it is assumed that the API has responded with information of a different entity; in this case, any matches found between the records must be ignored.

\subsection{Aligning Phase: Final Alignment}
Afterwards $A_{r_{in}}$ is used to determine the final alignment. For each relation in $A_{r_{in}}$ the best path match on the API side is searched (if existing). It is easy to match relations and paths without branching points, e.g. \texttt{label} or \texttt{title} in Figure~\ref{fig:paths} (see (1)). However, for matches with a branching point path (see (2)), we need to decide if all entries of the corresponding array provide the same type of information or different types (see Figure~\ref{fig:branching_example} for more details). In the first case, e.g. an array specifying the authors of a paper, we need to match \emph{all paths} that are equal (index values omitted) of the API response with the same relation. This is the case in the example in Figure~\ref{fig:request_response} for the path \texttt{authors.*.name}. In the last case, where every entry of the array has a different type, each different index value at the branching point should be mapped to one specific relation, possibly different relations for the different index values. In the example, \texttt{facets.0.value("2020")} always denotes the year of the publication, whereas \texttt{facets.1.value("Computer Science")} denotes the genre of the publication. Therefore, matching either the year or the genre relation of $K$ to \texttt{facets.*.value} is incorrect and should be prevented.

\begin{figure}[t]
\centering
\includegraphics[scale=0.62]{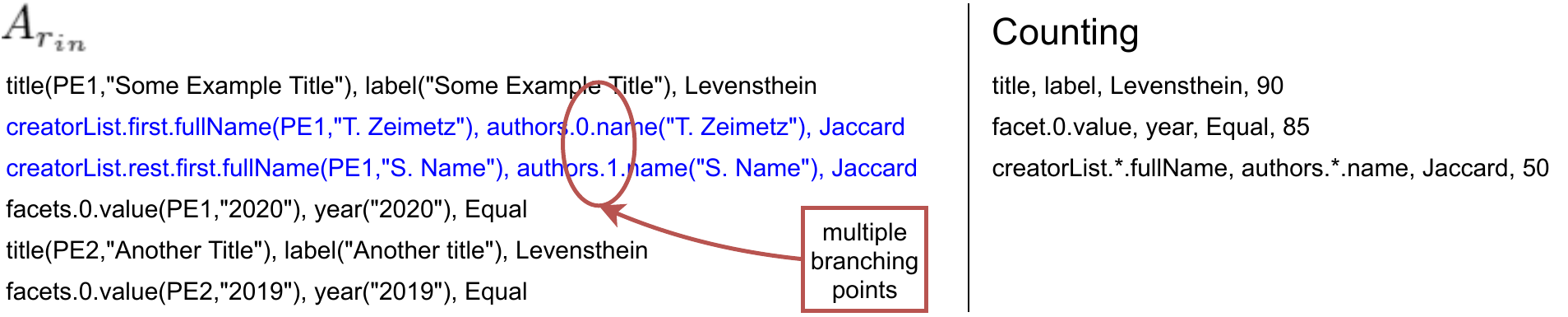}
\caption{Example Fragment of $A_{r_{in}}$ with abbreviated names for readability.}
\label{fig:alignment}
\end{figure}

In order to solve the problems above, \filipo\ distinguishes two cases: fixed path matches (FPM) and branching point matches (BPM). First, for every relation $r$ for which at least one tuple $(r,p,m)$ exists in $A_{r_{in}}$ we determine the path $P*$ (index values are replaced by the wild card symbol) that was matched most often in $A_{r_{in}}$, regardless of which similarity method $m$ was used.

Next it is determined if $(r,P*) \in A_{r_{in}}$ is a BPM or FPM. Hence, it is checked if the path $P*$ that was matched to $r$ in $ A_{r_{in}}$ only had one index value at the branching point or multiple different ones (see $A_{r_{in}}$ in Figure~\ref{fig:alignment}, blue highlighted lines). An example of a fixed path is \texttt{facets.0.value} in the set $A_{r_{in}}$ in Figure~\ref{fig:alignment}. To indicate the year, the same path is always used in the API response. The first entry of the array \texttt{facets} describes always the year of publication. If only one index value is found, it is considered as FPM. To ensure that it is a valid FPM, a confidence score for this match is determined. If a path was found only a few times, a match is not convincing. Hence, we calculate a confidence score for the matching by dividing the number of valid matches for $r$ by the number of responses. This confidence must be greater than the confidence threshold $\theta_{rec}$. We reuse $\theta_{rec}$ based on the assumption that the overlapping of records is also reflected in the overlapping of relations. In the example in Figure~\ref{fig:alignment} (right side) it is shown that for \texttt{facet.0.value} and \texttt{year} 85 matches (using Equal) are found. Assuming that 100 requests are sent to the API and all of them are answered, this results in a confidence score of $\frac{85}{100} = 0.85$ for this match. If the score is greater or equals than $\theta_{rec}$, it will yield as valid FPM and the relation-path match is added to the final alignment set. Note that another example for a FPM is the match of \texttt{title} and \texttt{label}.

Some relations and paths are dependent on the previous entity. For example, to match the name path for an author we have to include the whole author array of the API response because matching only one specific path (e.g. \texttt{authors.0.name}) excludes information of other authors. Hence, if more than one index value was found for $P*$ it is possible that $(r,P*)$ is a BPM. A match of $r$ and a branching point path $P*$ is considered valid if the following two conditions are satisfied: (1) if the relation $r$ has led to a match often enough, i.e. the previously computed confidence value is $\geq \theta_{rec}$, and (2) if the matched $(r,P*)$ occurs frequently enough in all matches $A_{r_{in}}$. If both conditions are met, the match $(r,P*)$ is considered a BPM and added to the final alignment set. In all other cases, the match was not convincing and is therefore discarded.

For the sake of simplicity, one aspect has yet not been considered in detail. Some relations can potentially be matched with multiple paths in the API response. For example, the relation for the publication year could be incorrectly matched with the path to the publication years of the article's references. To mitigate such errors, a reciprocal discount is used, i.e. the number $n$ of matches found for a possibly incorrect path $p$ and a relation $r$ is discounted by the length difference of the paths to $n/|(len(r)-len(p))|$. Thus paths with the same length (and potentially same structure) as the KB are preferred. At the end the final alignment set contains all valid matches found for the input relation $r_{in}$.

\section{Evaluation}\label{sec:evaluation}
\textbf{Baseline System.} Many systems~\cite{qian_2012,dhamankar_2004} in Section~\ref{sec:related_work} work semi-automatically with user assistance and are mostly designed for data of the same format. Some of the systems exploit schema information, use semantics or ``sameAs'' relations to find alignments. However, schema information exists rarely on the API side and using semantics or relations is difficult since API responses do not always have clear semantics. Furthermore, ``sameAs'' predicates are a concept of RDF and not present in classical API responses. Thus, we only use DORIS as baseline.

\textbf{Datasets and Platform.} We evaluated DORIS and \filipo\footnote{\url{https://github.com/dbis-trier-university/FiLiPo}} on three local KBs, seven bibliographic APIs and two movie APIs. One KB is an RDF version of the dblp\footnote{provided by dblp: \url{https://basilika.uni-trier.de/nextcloud/s/A92AbECHzmHiJRF}}. The other KBs are the Linked Movie DB\footnote{\url{http://www.cs.toronto.edu/~oktie/linkedmdb/linkedmdb-18-05-2009-dump.nt}} and a self created RDF version of IMDB\footnote{\url{https://www.imdb.com/}}, both containing movie information. The used APIs are SciGraph~\footnote{\url{https://scigraph.springernature.com/explorer/api/}}, CrossRef~\footnote{\url{https://www.crossref.org/services/metadata-delivery/rest-api/}}, Elsevier\footnote{\url{https://api.elsevier.com}}, ArXiv\footnote{\url{https://arxiv.org/help/api}}, two APIs provided by Semantic Scholar\footnote{\url{https://api.semanticscholar.org}} (one with DOIs and one with ArXiv keys as input parameters) and the COCI API of Open Citations\footnote{\url{https://opencitations.net/index/coci/api/v1}}. All of these APIs respond with metadata about scientific articles. To align the movie KBs we used the APIs of the Open Movie Database (OMDB)\footnote{\url{http://www.omdbapi.com}} and The Movie Database\footnote{\url{https://developers.themoviedb.org/3/find/find-by-id}}. It responds with metadata about movies, e.g. movie director and movie genres. All experiments are done on a workstation (AMD Ryzen 7 2700X, 48GB RAM) and all KBs are processed and stored as triple databases by using the Apache Jena Framework. However, note that the hardware used is not the bottleneck for the runtime, instead it is the latency and response speed of the APIs used.

As a gold standard\footnote{Code and gold standard can be found at~\url{https://zenodo.org/record/4778531}}, we manually determined the correct path alignments for each suitable combination of KB and API. Alignments were ignored that could not be determined based on the data, but for which a human may have been able to draw a connection. For example sameAs relations, in most cases, cannot be determined automatically since the URLs may differ completely.

{
\small
\begin{table*}[t]
    \caption{Probing Time (PT), Alignment Time (AT), (Average) Alignments (A), (Mean) Precision (P), (Mean) Recall (R), (Mean) F1 Score (F1)}
    \label{tab:filipo_eval}
    \centering
    \begin{tabu}{|l|c|c|c|c|c|c|c|c|c|c|c|}\hline
        &  \multicolumn{7}{c|}{\textbf{FiLiPo}} & \multicolumn{4}{c|}{\textbf{DORIS}}\\
        Data Sets & Requests & APT & AT & A & P & R & F1 & A & P & R & F1 \\
        \hline
        dblp $\leftrightarrow$ CrossRef & 25/75 & 18.0 & 4.0 & 18 & 0.97 & 0.78 & \textbf{0.91} & 9 & 0.89 & 0.36 & 0.51\\
        dblp $\leftrightarrow$ SciGraph & 25/75 & 14.5 & 2.5 & 18 & 0.96 & 0.78 & \textbf{0.86} & 11 & 1.00 & 0.38 & 0.55\\
        dblp $\leftrightarrow$ S2 (DOI) & 25/75 & 24.5 & 8.0 & 15 & 0.89 & 0.87 & \textbf{0.88} & 12 & 0.83 & 0.47 & 0.60\\
        dblp $\leftrightarrow$ S2 (ArXiv) & 25/75 & 24.5 & 9.0 & 7 & 1.00 & 0.88 & \textbf{0.94} & 6 & 0.83 & 0.33 & 0.47\\
        dblp $\leftrightarrow$ COCI & 25/75 & 23.0 & 19.0 & 16 & 1.00 & 0.78 & \textbf{0.88} & 9 & 1.00 & 0.33 & 0.50\\
        dblp $\leftrightarrow$ Elsevier & 25/375 & 17.5 & 5.5 & 13 & 0.92 & 0.92 & \textbf{0.92} & 13 & 0.85 & 0.85 & 0.85\\
        LMDB $\leftrightarrow$ TMDB & 25/75 & 4.5 & 2.0 & 6 & 0.94 & 1.00 & \textbf{0.97} & 7 & 0.57 & 0.80 & 0.67\\
        \hline
        dblp $\leftrightarrow$ ArXiv & 25/75 & 11.5 & 3.5 & 8 & 0.83 & 0.86 & \textbf{0.85} & 5 & 1.00 & 0.43 & 0.60\\
        LMDB $\leftrightarrow$ OMDB & 25/75 & 36.5 & 3.5 & 14 & 0.93 & 0.95 & \textbf{0.94} & 11 & 0.55 & 0.56 & 0.55\\
        IMDB $\leftrightarrow$ OMDB & 25/75 & 4.0 & 14.5 & 9 & 0.73 & 0.66 & 0.69 & 9 & 1.00 & 0.90 & \textbf{0.95}\\
        \hline
    \end{tabu}
\end{table*}
}

\textbf{\filipo\ Evaluation.} In order to find a suitable configuration (sample size and similarity thresholds) that works for most APIs, we performed several experiments. \filipo\ works with two different thresholds: the string similarity $\theta_{str}$ and the record overlap $\theta_{rec}$. To determine a combination of both thresholds that provides good alignment results, we tested several combinations of values for both thresholds (steps of 0.1) and calculated precision, recall and F1 score. The found alignments had a very high precision for $\theta_{str}$ between 1.0 and 0.5; recall was significantly better at 0.5. This is mainly due to the fact that data which are slightly different (e.g. names) can still be matched. For large values of $\theta_{rec}$, many alignments are lost, because the data of a KB and an API overlap only slightly in the worst case. Here, a value of 0.1 to 0.2 was already sufficient to prevent erroneous matching. Hence, we  used $\theta_{str}=0.5$ and $\theta_{rec}0.1$ in the experiments. Regarding the sample size we determined that 25 probing and 75 additional requests (sample size of 100) is suitable for most APIs. However, since some KBs and APIs have little data in common, the sample size may need to be adjusted.

We assume that users have no in-depth knowledge of used APIs, but are familiar with the structure of the KB and that users have domain knowledge and hence understand common data structures from the genre of the KB (e.g. bibliographic meta data). Additionally, users can make further settings (e.g., changing the sample size) to fine-tune \filipo. All APIs were executed with default settings, i.e. 25 probing with 75 additional requests (in total 100 requests) are made for every valid input relation. Since dblp contains relatively few publications by Elsevier, we set the number of additional requests for Elsevier to 375.

One current limitation of \filipo\ becomes clear when using IMDB. We had to set the record overlap threshold from 0.1 to 0.3. This is because IMDB contains several relations with low functionality (e.g. \texttt{movieLanguage}) and hence incorrect matches would be tolerated. In contrast, DORIS excludes all relations with a very low functionality from the alignment process. Hence, it prevents the result for erroneous matches but but also loses some (possibly important) matches.

Since \filipo\ pulls random records from a KB and uses them to request an API, the alignments found may differ slightly between different runs. Hence, the evaluation was performed three times for each combination of KB and API. The average runtime was approx. 25 minutes. If input relations are known, as is the case with DORIS, then the system usually needs no longer than a few minutes because the probing phase can be skipped. The probing phase is expensive in runtime because an API is requested a significant number of times (see Table~\ref{tab:filipo_eval}).

\filipo\ was able to identify all correct input relations for almost all APIs. The only exception is the combination of IMDB and OMDB: IMDB contains a relation (\texttt{alternativeVersion}) to specify an alternative version of a movie  (e.g. a directors cut is an alternative version of a movie) which is a valid input for OMDB. Of four possible input relations, it was able to identify all (alternative) title relations (\texttt{title}, \texttt{label} and \texttt{alternativeTitle}) as input relation in all runs, but only determined in 60\% of the runs the \texttt{alternativeVersion} as input. The reason for this is that especially the alternative version of lesser-known movies are unknown to OMDB. It can be summarised that \emph{in all cases a valid input was found} but only in 9 of 10 cases \emph{all} valid input relations were found.

For the evaluation of the alignments we used the metrics precision, recall and F1 Score. \filipo\ was able to achieve a precision between 0.73 to 1.00 and a recall between 0.66 to 1.00. Values close to 1.0 were achieved mainly because there were only a few possible alignments. The corresponding F1 scores for \filipo\ are between 0.69 and 0.95.

\textbf{Baseline Evaluation.} We re-implemented DORIS for our evaluation. It uses label information of instances as its predefined input relation for APIs. Since this is not always the appropriate input parameter for an API (e.g. some APIs expect DOIs as input) we modified DORIS such that the input relation can be specified by the user. Compared to \filipo, DORIS has an advantage in the evaluation, since it does not have to determine valid input relations for the used APIs. In contrast to \filipo, these input relations must be specified by the user and hence the runtime is shorter and there is no risk of alignment with wrong input relations. DORIS uses two different confidence metrics to determine an alignment: the overlap and PCA confidence. We assessed that the PCA confidence delivers better results for the alignment and hence DORIS is able to match journal-related relations. Since most of the entities in dblp are conference papers, journal specific relations are lost when using the overlap confidence. The downside is that a path that was found only once in the responses only needs to match once to achieve a high confidence. In such cases it is risky to trust the match and hence a re-probing is performed which increases the runtime considerably, since entities that share the matched relation are subsequently searched and ranked. DORIS has been configured in order to send 100 requests to the APIs. The threshold for the PCA confidence has been set to 0.1 based on a calibration experiment similarly to \filipo\, testing several threshold values between 0.1 and 1.0 (in steps of 0.1). With threshold 0.1, no erroneous alignments were made; recall was significantly larger at 0.1 than with larger values. 

\filipo\/ outperforms DORIS in terms of precision in most cases and clearly in terms of recall and F1. This is mainly caused by the two disadvantages of DORIS discussed before: First, aligning with entities with most facts often misses rare features of entities (e.g. a specific publisher like Elsevier). As a result, it is not possible for DORIS to determine an alignment between dblp and Elsevier's API. Second, using only one similarity method results in a relatively high precision, but is also too rigid to recognise slightly different data (abbreviations of author names), thus leading to low recall. However, DORIS was able to achieve better results using IMDB, mainly because DORIS excludes all relations with a very low functionality from the alignment process. This is also the reason why DORIS was significantly worse in terms of recall in the other alignment tests. However, since OMDB responds with only a small amount of information, in which no information with a high functionality was included, this limitation does not have a negative effect but rather a positive one. 

\section{Conclusion}
We presented \filipo, a system to automatically discover alignments between KBs and APIs. A user only needs knowledge about the KB but no prior knowledge about the APIs data schema. In contrast to the current state-of-the-art system DORIS, our system is additionally able to determine valid input relations for APIs which significantly reduces manual effort by the user. In all cases a valid input relation was found by \filipo\ but only in 9 of 10 cases \emph{all} input relations were found. Our evaluation showed that \filipo\ outperformed DORIS and delivered better alignment results in all but one case. As \filipo\ is currently only able to determine 1:1 matches, we are already working on a version that will also allow 1:n matches.

\bibliographystyle{splncs04}
\bibliography{database}

\end{document}